\documentclass[aps,prb,twocolumn,showpacs,floatfix]{revtex4}
\usepackage{graphicx}

\begin{document}

\title{Orbital order in NaTiO$_2$ : A first principles study}
\author{Monika Dhariwal$^{1}$, T. Maitra$^{1,*}$, Ishwar Singh$^{1}$, 
S. Koley$^{2}$ and A. Taraphder$^{2}$}
\affiliation{$^{1}$Department of Physics, Indian Institute of Technology,  
Roorkee - 247667, Uttarakhand, India}
\affiliation{$^{2}$Department of Physics and Centre for Theoretical Studies, 
Indian Institute of Technology, Kharagpur - 721 302, India}
\pacs {71.20.-b, 71.15.Mb, 75.25.Dk}                                                                                                   
\date{\today}
\email [corresponding author:]{tulimfph@iitr.ernet.in}

\begin{abstract}
The debate over the orbital order in layered triangular lattice system 
NaTiO$_2$ has been rekindled by the recent experiments of McQueen, et al.
\cite{cava} on NaVO$_2$ ({\em Phys. Rev. Lett.} {\bf 101}, 166402 (2008)). 
In view of this, the nature of orbital ordering, in both high and low 
temperature states, is studied using an ab-initio electronic structure 
calculation. The orbital order observed in our calculations in the low 
temperature structure of NaTiO$_2$ is consistent with the predictions of 
McQueen, et. al. An LDA plus dynamical mean-field calculation shows 
considerable transfer of spectral weight from the Fermi level but no metal-
insulator transition, confirming the poor metallic behaviour observed in 
transport measurements.      
\end{abstract}
\vspace{0.5cm} 
\maketitle

Transition metal ions sitting at the vertices of triangles in a layered 
triangular lattice system present a very interesting playground for the study 
of competing interactions\cite{takada,schaak,giot}. The presence of unfilled 
d-levels in the transition metal ion brings in correlations while the 
geometric frustration of the underlying lattice tends to prevent long range 
antiferromagnetic (AF) order. Consequent large entropy in the ground state 
usually relieves itself via broken symmetry at lower temperatures\cite{cava,
clarke}. If the transition metal ion is Jahn-Teller active, it is likely to 
induce frozen lattice distortions. It has been observed that orbitals often 
play an active role in removing the frustration leading to an orbitally 
ordered (OO) ground state\cite{cava,pen}.  

The ongoing debate~\cite{oles,pen,ezhov} about the role of orbitals 
in quenching the spin requires more careful and diverse work. There is only a 
limited experimental and theoretical work available in the literature
on NaTiO$_2$ to take a definitive call on this. 
It is already emphasized in earlier theoretical work~\cite{ezhov,pen,khom}  
that the spin degree is quenched via an orbitally induced Peierls transition 
at around 50K and the low temperature physics is dominated by the orbital 
correlations.  

The absence of magnetic LRO, very small susceptibility below 250K, lack of 
Curie-Weiss behavior in any region of temperature, apparent Pauli 
paramagnetism above and below the broad transition region - virtually all 
experimental evidence so far, point towards a spin degree which is completely 
quenched at low temperature~\cite{clarke}. In addition, the entropy loss much 
in excess of $Rln2$ at the transition is a strong indication of the 
involvement of orbital degrees in the ordering process. While the transition
is likely to be affected by spin fluctuations (getting quenched there -  
indicated by the drop in spin susceptibility), the role of spin degrees in 
selecting the ground state order is not clear. In fact most of the model 
calculations appear to support an orbitally driven symmetry breaking in the 
ground state. 

There is a structural transition around 250K in NaTiO$_2$ with four short 
and two long Ti-O bonds per octahedron in the low temperature phase~
\cite{clarke}. Orbital order, concomitant with the structural transition, has 
been reported recently in NaVO$_2$~\cite{cava} from high resolution XRD and 
powder neutron diffraction (PND). Both in structure and electronic 
configuration, NaTiO$_2$ is very similar to NaVO$_2$, where a transition driven 
by OO has been proposed earlier~\cite{ezhov,pen,cava}.  
However, it is not established yet what exactly is the nature of OO in 
the low temperature phase. Indeed, being a d$^{1}$,\, $s=\frac{1}{2}$ system 
without a magnetic long range order (LRO), NaTiO$_2$ was hotly pursued 
originally as a candidate for spin liquid~\cite{cava}. 

The structural transition that NaTiO$_2$ undergoes is from the high 
temperature rhombohedral phase (space group $R\bar{3}m$) to a monoclinic phase 
(space group $C2/m$) through a broad, continuous transition between the 
temperature range 220K to 250K\cite{clarke}. In the rhombohedral phase 
Ti ions form  2D triangular planes separated by alternating Na and O 
planes (arranged as Na-O-Ti-O-Na$\cdots$ planes) along [111] direction. The 
interlayer Ti-Ti distance is much higher than intralayer Ti-Ti distance. It is 
therefore generally assumed that the interaction among the Ti ions of different layers is not relevant. Furthermore, the Ti ions are octahedrally coordinated 
by O ions and 
these TiO$_6$ octahedra share edges with each other. In the high 
temperature rhombohedral phase all six oxygens surrounding a Ti ion are equally distant and 
the Ti ions form isosceles triangles with all the Ti -Ti bonds being equal
inside a layer. On cooling, it aquires a monoclinic structure where  
two among the six Ti-O bonds 
elongate and the remaining four shorten affecting Ti -Ti 
bonds as well: out of the six nearest neighbour intralayer Ti-Ti bonds, four 
are shortened and two elongated. This structural transition should evidently  
affect the electronic structure. 

Previous electronic structure calculations~\cite{ezhov} within LDA+U approach 
predicted an insulating state for the high temperature rhombohedral phase. 
Whereas, experimentally, the system remains metallic, albeit with high 
resistivity, at all temperatures~\cite{clarke}. There are, however, no 
electronic structure calculation available in the literature for the low 
tempretaure phase. In this context, therefore, a first principles calculation 
is quite pertinent. We investigate in detail the electronic structure of both 
high and low temperature phases from first principle calculations using 
linearized augmented plane wave (LAPW) method as implemented in the Wien2k 
code~\cite{wien2k}. 

We have taken the crystal structure and atomic positions from Clarke 
et al.~\cite{clarke}. 
In the high temperature rhombohedral phase, as already discussed, 
all Ti-O bonds in a TiO$_6$ octahedron are of same length, 2.08\,\AA, \, 
but there exists a trigonal distortion making O-Ti-O angles different 
from 90$^{\circ}$. All intralayer Ti-Ti nearest neighbour bonds are of 
equal length as well, 3.04~\AA. As discussed above, in the low temperature 
monoclinic structure two of the six Ti-O bonds in a TiO$_6$ octahedron are 
stretched to 2.11 \AA \, while the remaining four Ti-O bonds become 2.05 \AA \, each. However, the  O-Ti-O bond angles get closer to 90$^{\circ}$ in this 
phase. Similarly six Ti-Ti bonds also 
divide into four short bonds (3.028\AA) and two long bonds (3.03 \AA). So the 
Ti-Ti bonding increases in the low temperature phases which is reflected in 
the increase of conduction bandwidth by about 0.2 eV. Note that the distortion 
in Ti-O bonds is much higher than that of Ti-Ti bonds.   
In our LAPW calculation, the muffin tin sphere radii for Na, Ti and O were 
chosen to be 2.22, 2.05 and 1.82 a.u. respectively and approximately 200 $k$-
points were used in the irreducible first Brillouin zone for the calculation. 

\begin{figure}
\includegraphics[width=8cm]{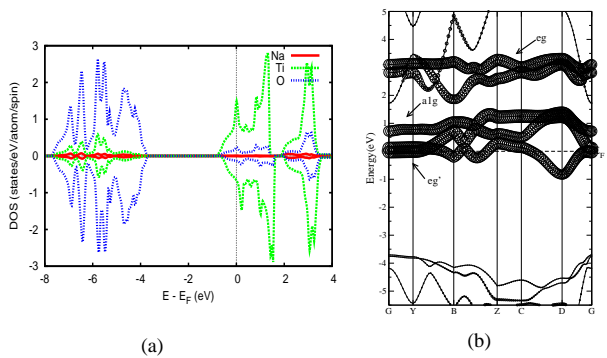}
\caption{(color online) Spin polarised density of states and LDA band structure 
(spin up only) of NaTiO$_2$ in the low temperature monoclinic phase. 
Bands are plotted in high symmetry directions and the d-orbitals are shown in 
the usual fat band scheme.} 
\end{figure}

We present in Fig.1 the density of states (DOS, in both spin directions) and 
the band structure (for spin up) in certain high symmetry directions 
for the low temperature monoclinic structure within local spin density 
approximation (LSDA). The $d$-bands are shown in the usual fat band scheme 
- where the width of the band represents the contribution of d-orbitals to 
the respective band, revealing the orbital contributions along
different symmetry directions. The $e_g$-$t_{2g}$ splitting 
due to the octahedral crystal field of oxygens is clearly visible. Presence of 
trigonal distortion, described above, further splits the $t_{2g}$ levels into
an $e_g^{\prime}$ doublet and an $a_{1g}$ singlet. The shape of $a_{1g}$ 
singlet  and $e_g^{\prime}$ doublet orbitals are shown in Fig. 2(a) and (b) 
respectively. The $e_g^{\prime}$ doublet is found to be lower in energy than 
the $a_{1g}$ singlet, in contradiction to the predictions of crystal field 
theory as explained below. 

This feature has been reported recently by Jia et. 
al\cite{jia} for the isostructural system NaVO$_2$ and has also appeared~\cite{Co} in 
the study of Na$_x$CoO$_2$. The 
argument presented in all these cases is that there is a competition between 
crystal field effects and e$_g$ - $e_g^{\prime}$ hybridization. Discussed at 
length by Landron and Lepatit \cite{landron}, it is the hybridization between 
$e_g^{\prime}$  and e$_g$ orbitals (owing to their similar symmetries) which 
plays a dominant role in deciding the 
relative order of $a_{1g}$ and $e_g^{\prime}$ orbitals which could often go 
against the crystal field theory predictions. 
These authors have shown from quantum chemical calculations in clusters,  
that interations like metal-ligand hybridization, long range crystalline 
field effects, screening etc. have negligible effect in this regard. 
They have further shown that the value of the angle
$\theta$ between three-fold (111) direction of the MO$_6$ octahedron and 
the M-O direction (schematically shown in Fig. 2(c) for TiO$_6$)
plays a crucial role in dictating the energetics of a$_{1g}$, $e_g^{\prime}$
splitting. For a regular octahedron this angle is calculated to be 54.74 
degrees~\cite{landron} while 
from the experimental structure data~\cite{clarke} we estimate the angles 
to be 57.78$^\circ$ and 56.54$^\circ$ in the high and low temperature 
phases of NaTiO$2$ respectively. So there is a (slight) compression along 
the (111) direction in comparison to the regular octahedron. This compression
is also present in other isostructural systems such as NaVO$_2$, Na$_x$CoO$_2$. Under this trigonal compression, crystal field theory predicts $a_{1g}$ orbital to be lower and  $e_g^{\prime}$ 
orbitals to be higher in energy. However, we observe the opposite splitting 
like in the cases of NaVO$_2$ and Na$_x$CoO$_2$. Therefore, we believe that
$e_g$ - $e_g^{\prime}$ hybridization plays a dominant role in deciding
the $t_{2g}$ splitting in NaTiO$_2$ as well. 
\begin{figure}
\includegraphics[width=8cm]{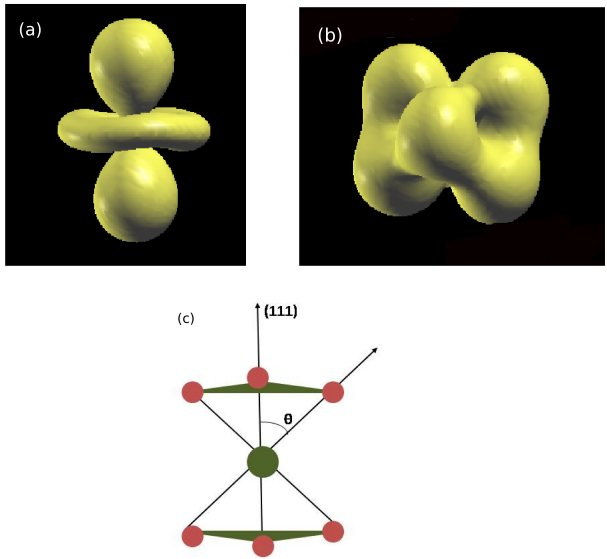}
\caption{(color online) Shape of (a) a$_{1g}$ (singlet) and (b) e$_g^\prime$ 
(doublet) orbitals formed after the splitting of t$_{2g}$ orbitals under the 
trigonal distortion.  
(c) The angle $\theta$ measuring trigonal compression of the octahedron.}
\end{figure}

In the presence of transition metal ion Ti, with partially filled 3d-shell, 
the correlation effects are bound to be important. Furthermore,
in order to ascertain the presence of any orbital order 
one needs to incorporate the Coulomb correlation as it stabilizes the order
further. Therefore, we undertake an LSDA+U calculation and discuss our results below. 
From our calculations (with U$_{eff}$ = 3.6 eV \cite{ezhov}) we observe that 
there is a clear splitting between $a_{1g}$ and $e_g^{\prime}$ now (which was 
quite small in LSDA) and the relative order is indeed the same as in LSDA. 
We present the LSDA+U DOS (partial DOS for d-orbitals) and band structure (spin
up direction only) in Fig. 3. It is also seen that 
on applying $U$ all the down-spin weight of the $d$-states just below the 
Fermi level gets transferred to the spin-up sector leaving spin-down sector 
empty. The system appears to be half-metallic with 100\% spin polarization 
at the Fermi level.  

We have calculated the electron density at each Ti site to study if there is 
any orbital order and in case there is one, then the nature of the order in the low 
temperature phase of NaTiO$_2$. In 
Fig. 4(a) we show the real space electron density plot of the 
orbitals occupied by the single d-electron at each Ti site. Comparing the 
electron density at each Ti site with that of Fig. 2(b) (where the shape of
the $e_g^{\prime}$ doublet is shown), one can clearly see that both the 
$e_g^{\prime}$ orbitals are occupied by the single d-electron of Ti ion at 
each site. Note that in Fig. 2(b) the 3D orbitals are shown from one side 
whereas in Fig. 4(a) they are seen from the top with a slight tilt so 
they might look a bit different but they have the same 3D structure.  
\begin{figure}
\includegraphics[width=8.5cm]{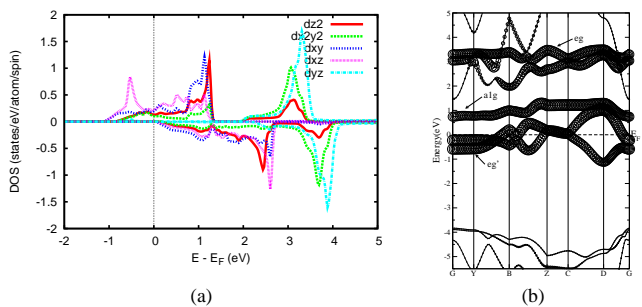}
\caption{(color online) Partial DOS for five d-orbitals around the Fermi 
level in both 
spin directions are shown in (a). Fermi level falls only in the spin up DOS.
In (b) the bands are shown only for spin up direction in the fat band 
scheme highlighting the five d-bands. } 
\end{figure}

Looking at Fig. 4(a) more closely, it can also be seen that the lobes of the 
occupied $e_g^{\prime}$-orbitals at each Ti site are oriented in such a 
way so as 
to have an anisotropy in the interactions (as explained below) with the six 
nearest neighbour Ti ions in the hexagonal plane. Along a particular direction
(indicated here by thick lines) there would be comparatively more interaction 
among the neighbouring Ti d-orbitals because lobes along this direction
face each other more directly than along the other directions (shown by the 
thin 
lines). Orbital lobes facing each other along the thick line are indicated by
curves with double ended arrows in Fig. 4(a) (with solid curves pointing at 
lobes which are above the hexagonal plane and dashed curves pointing at lobes 
which are below). Note that the lobes we are talking about here 
do not lie in the same haxagonal plane formed by the Ti ions, they lie in a 
plane which is nearly perpendicular (with a slight tilt) to the former. 
Therefore, the orbital overlap among the neighbouring Ti ions along 
this direction is not of ``in-plane'' $\sigma$-
type, but it is ``out of plane`` $\pi$-type in which two lobes bind above the hexagonal plane
and two lobes bind below. To make the above picture clearer we present in 
Fig. 4(b) a 2D intersection of the electron density in a plane (the red 
plane in Fig. 4(b)) which is nearly perpendicular to 
the hexagonal Ti plane. One can easily see that lobes of 
the xz-type orbitals at neighbouring Ti sites lie in the same plane (the red plane) along 
z-direction (thick 
lines in Fig. 4(a)) whereas they lie in planes parallel to each other along the 
y-direction. In this figure we have only shown 2D electron densities for some 
selected Ti ions for demostration purposes. 

From the above discussion it is
clear that there exists some anisotropy in the interactions among nearest 
neighbour (nn) Ti ions along six different directions due to the directional 
nature of occupied orbitals at these sites. Among the six nn interactions,
two (indicated by the thick lines in Fig. 4(a)) are more interacting than the 
other four (indicated by the thin lines in Fig. 4(a)). This is also reflected 
in the distortion of Ti-Ti bond distances observed experimentally in the low
temperature phase as explained in the following. Due to the Coulomb repulsion
being more along the two directions than the remaining four directions, the Ti-
Ti bond distances are slightly higher (3.03 \AA) than the remaining four Ti-Ti 
distance (3.028 \AA). Thick lines in Fig. 4(a) correspond to higher bond 
distances and thin ones indicate lower bond distances.

\begin{figure}
\includegraphics[width=8cm]{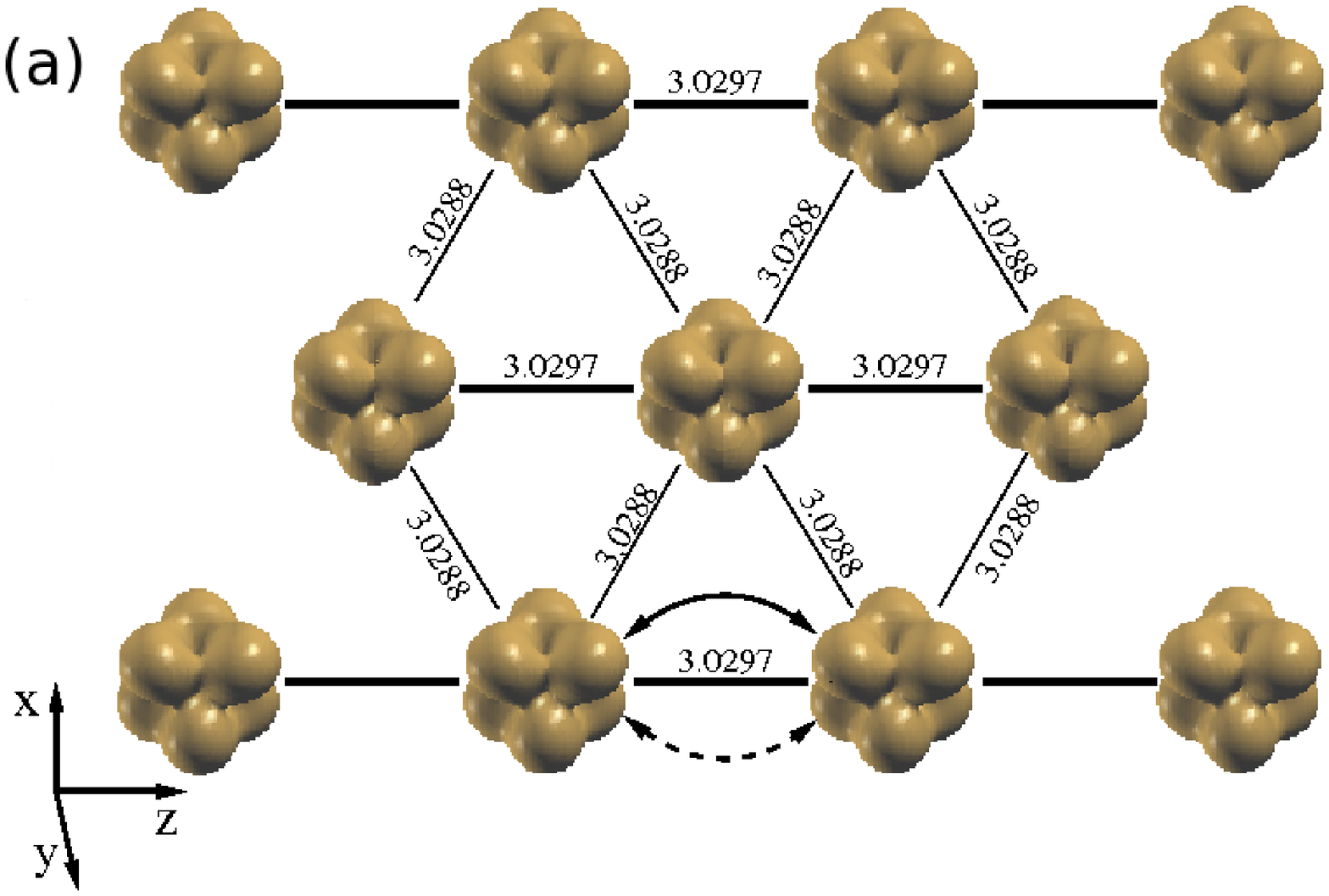}
\includegraphics[width=6cm]{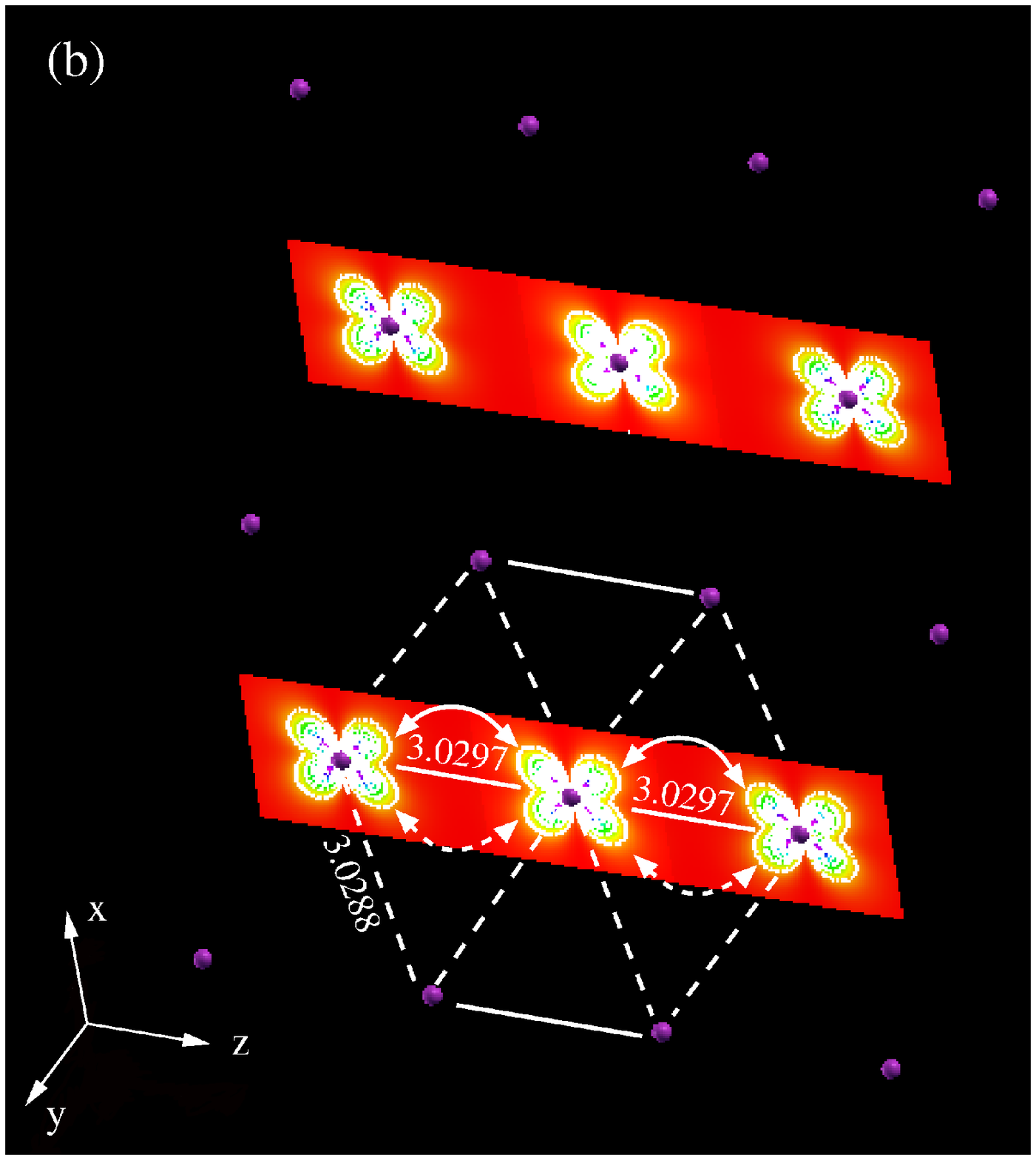}
\includegraphics[width=7cm]{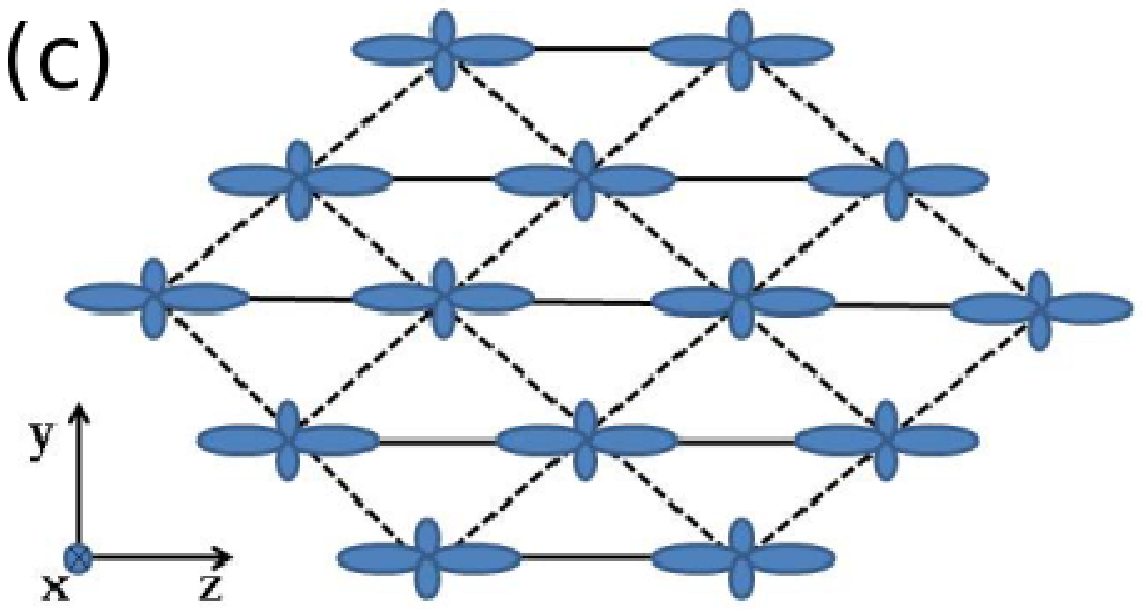}
\caption{(color online) (a) Electron density plot showing chain like orbital
order in the hexagonal Ti plane (thick black lines indicate the orbital
chain directions). Curved double ended arrows point towards the orbital lobes
which participate in $\pi$ type bonding along the orbital chains (see text), 
(b) 2D intersection of electron density in a plane (red plane) which is nearly 
perpendicular to the hexagonal plane formed by the Ti ions (small violet 
spheres indicate Ti ions). Dashed straight lines represent the shorter bonds 
and solid straight lines represent the longer bonds in the hexagon drawn. 
The two representative red 
planes shown are parallel to each other and each of them is nearly perpendicular to the hexagonal Ti plane.
(c) Schematic view of the orbital
order in the hexagonal plane formed by Ti ions as seen from the top. Dashed line
represent the shorter bonds whereas the solid lines represent the longer bonds.
Note that interaction along the solid lines (orbital chains) is of 
$\pi$-type and it is the plane projection of the orbital lobes from 
Fig. 4(a) that makes it appear like $\sigma$-type.}
\end{figure}
 
Following the discussion above, one can infer that the occupied orbitals at Ti
sites appear to form  one dimensional orbital chains in the hexagonally 
arranged Ti plane. The thick lines in Fig. 4(a) indicate 
the orbital
chain direction. Along the chain, a {\it ferro}-orbital order is observed with 
the lobes of the orbitals oriented in such a way (as shown above) 
so as to give rise to slightly stronger interaction among nearest neighbour 
Ti ions compared to that among the neighbouring chains. A schematic diagram 
with  the orbital 
order proposed above is shown in Fig. 4(c) where the inplane projection on the 
hexagonal Ti-plane is shown. Note that the schematic Fig. 4(c) does not 
indicate sigma 
bonding along solid lines, it is a 2D projection from Fig. 4(a) that makes it 
appear so.
 
McQueen, et. al.\cite{cava} predicted a similar orbital order for NaTiO$_2$ 
using physical arguments, based on their experimental observations 
on NaVO$_2$.  
This orbital ordering in the low temperature phase is a clear manifestation
of the structural transition from rhombohedral to monoclinic which makes two
out of the six Ti-O bonds in an octahedron unequal. So a reasonable
distortion in the TiO$_6$ octahedra gives rise to a small but finite distortion
in the inplane Ti-Ti bonding which in turn results in chainlike orbital order.
This orbital order is found to remain stable in the range of $U$ values 
varying from 2 to 7eV. 

\begin{figure}
\includegraphics[width=0.7\columnwidth]{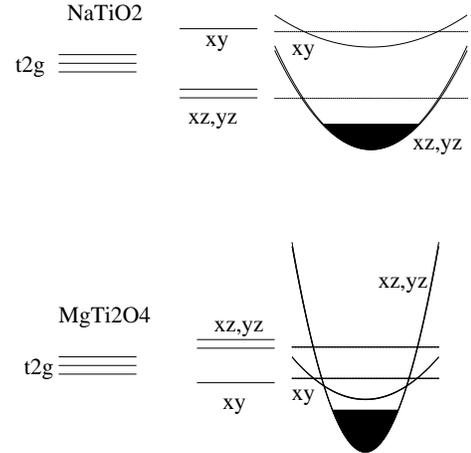}  
\caption{(color online) Schematic diagram comparing the electronic structure of 
NaTiO$_2$ and MgTi$_2O_4$ in their low temperature phases. The parabolas on the right represent xy or xz/yz
bands. The dashed lines show the reference positions of the centre of the
band. } 
\end{figure}
\begin{figure}
\includegraphics[angle=270,width=0.7\columnwidth]{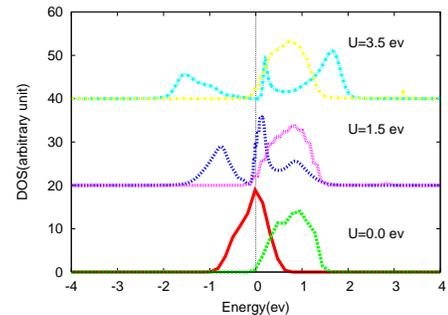}  
\caption{(color online) LDA+DMFT DOS of the valence and conduction bands 
of NaTiO$_2$ in the low temperature phase for two different $U$ values. } 
\end{figure}

We have also looked at the electronic structure of the high temperature 
rhombohedral phase using LSDA and LSDA+U methods. We do not observe any 
significant difference in the band structure from that of the low temperature
phase. With LSDA+U we get a half metallic state rather than an insulating state 
as observed by Ezhov et. al.\cite{ezhov}. Note that Clarke et al. predicted, 
from their experimental measurement, that there should not be any significant  
difference in the band structures across the structural transition. This is  
borne out in our DFT calculations too. Furthermore, their measurements also 
show that the system never becomes an insulator at any temperature, though the 
resistivity in the low temperature phase is fairly high, indicative 
of a ``bad metal."

The apparent similarity with the spinel MgTi$_2$O$_4$ prompted
Khomskii and Mizokawa\cite{khom} to speculate that
the orbital order of NaTiO$_2$ is similar to MgTi$_2$O$_4$.
However, we find that in spite
of their apparent similarity,  there are certain important differences between
the two systems. Due to the presence of strong trigonal
distortion as well as tetragonal elongation, the t$_{2g}$ band splits in the
opposite way to that of MgTi$_2$O$_4$ as shown in the schematic picture
comparing the electronic structures of the two in Fig. 5. Secondly due to the
two dimensional nature of Ti-layer in NaTiO$_2$ the t$_{2g}$ orbitals do not
point towards each other (which would have given $\sigma$-bonding) and hence 
the overlap is not as strong as in the spinel MgTi$_2$O$_4$. Therefore the
bandwidths of the xy or xz/yz orbitals do not significantly
increase due to the tetragonal distortion unlike in the case of spinels
\cite{khom}. With dynamical shifting of spectral weight (see below) this band
therefore remains more susceptible to the effects of strong Coulomg
interaction, making the system a poor metal. Whereas, in case of MgTi$_2$O$_4$,
due to the significant increase in the band width, xy/yz doubly degenerate
band extends below the singlet xy band and gets occupied by the Ti d-electron
(see Fig. 5 (lower panel)).

Though the orbital polarization of both NaTiO$_2$ and MgTi$_2$O$_4$ in their
low temperature phases looks similar from the band structure calculations
(i.e. both xz and yz orbitals are being occupied), the orbital order we find in 
NaTiO$_2$ is different: there is no indication of the dimerization of orbitals
(i.e., $yz-yz-zx-zx-yz-yz$). We rather find that both the
orbitals are occupied at each site with their lobes oriented in such a way
(as explained above) that one dimensional orbital chains are formed along
a particular direction in the Ti plane. Over a physically reasonable range of
U values (2-7 eV) we find this orbital order to remain stable and the system
also remains metallic.
Our observation of chain-like orbital order appears to be compatible with
the scenario predicted in
Ref\cite{cava} and Ref. \cite{pen}. Though this type of orbital order
would prefer an antiferromagnetic exchange interaction among the ferro-orbital
chains (according to Goodenough-Kanamori-Anderson rule\cite{gka}),
the available neutron diffraction measurements \cite{clarke} do not indicate
any long or short range magnetic order in this systems.

Although LDA+U method gets qualitatively right broken symmetry ground states, 
it is well known that the spectral weight transfer (SWT) due to correlations is 
beyond such static approaches. They also overestimate the `gap' in the DOS.  
Besides, the nature of the low temperature `bad metallic' state is beyond the
scope of LDA or LDA+U calculations. The SWT is at the heart of the Mott physics
and therefore we undertake an LDA plus multi-orbital dynamical mean-field 
theory
(MO-DMFT) calculation using the two relevant bands (of e$_{g}^{\prime}$ symmetry) 
close to the Fermi level. 
DMFT is exact in the infinite dimension and has been very successful in predicting 
the metal-insulator transitions and photoemission data in a host of correlated
systems~\cite{dmft}. It captures the dynamical correlations quite well 
while mapping the problem onto an impurity model. We use the iterated perturbation 
theory to solve the corresponding impurity model. 

Shown in Fig. 6, with increasing intraband correlation $U$ (taken the same for the
two bands), are the gradual evolution of the DOS for the two bands. Two features are
prominent here: (i) there is a gradual erosion of states at the Fermi level as $U$
increases and (ii) SWT over a large range of energy that can be identified in 
photoemission spectroscopy. The erosion of DOS leads to a pseudogap like structure
at the Fermi level consistent with the high resistivity observed. Even at a reasonably 
higher $U$, we do not observe a real gap opening at the Fermi level and the system
remains nominally metallic. A clear signature of the building correlation is the 
appearence of the lower and upper ``Hubbard bands" in the spectral density. 
The lower Hubbard band at around -1 to -2 eV can be easily  picked up in 
a photoemission experiment confirming the correlated nature of the bands.  
We also note that the effect of an interband correlation 
is very small in the present case as the Fermi level is almost entirely in 
the lower band. Depending on the orbital characters of the two bands, this may 
indicate the possibility of preferential orbital occupation 
(``orbital selective" metallic state) and an orbital ordering. However, 
we do not see a Mott transition upto $U=4.5$eV.  

Finally, we sum up our findings from this first ab-initio calculation for the 
electronic structure and orbital order of the low temperature monoclinic phase 
of NaTiO$_2$. As we have mentioned above, not much experimental or theoretical
works are available to compare our results with. However, recent experimental
measurements in the isostructural compound NaVO$_2$
\cite{cava} have been used to make some predictions about the possible orbital 
order in NaTiO$_2$. Our results  appear to bear out their suggestions and 
are consistent with the possible order predicted thereof. Our LDA calculations 
reveal metallic states both above and below the structural transition which
is consistent with the experimental observations as well. However, the metallic
state found experimentally is a 'bad metal' while the LDA DOS at Fermi level
is fairly high. We also find that the LDA+U does not change this, though it 
stabilizes the orbital order further. This implies that the 'bad metal' 
state requires inclusion of dynamical fluctuations beyond single-particle 
description. Finally, incorporating dynamical corrections using   
LDA+DMFT calculations we show how large spectral weight transfer across the
Fermi level leads to a bad metallic state as observed experimentally. 
We also observe that the orbital order of NaTiO$_2$, speculated to be similar 
to that of MgTi$_2$O$_4$ (a Ti-spinel compound)\cite{khom}, has certain 
differences, though the orbital polarization is same in both the systems. 
The dimerization of orbitals forming singlet pairs, observed in MgTi$_2$O$_4$, 
does not occur in NaTiO$_2$.  

In conclusion, we have studied the electronic structure of NaTiO$_2$ in both 
high temperature rhombohedral phase and low temperature monoclinic phase. 
We provide a theoretical resolution that the low symmetry state is, in fact, 
a poor metal and not an insulator. We also find a considerable spectral 
weight transfer across the range and make predictions for the spectral 
densities that can be seen in photoemission experiments to establish the 
correlated nature of electronic states. Furthermore, in the low 
temperature phase, the system is found to be orbitally ordered with one 
dimensional chain-like order along a particular direction in 
the hexagonally aranged Ti-plane. The orbital order found here is 
consistent with the recent predictions made by McQueen et. al.\cite{cava} on 
the basis of their experimental observations on a similar system NaVO$_2$. 

\noindent{\bf Acknowledgement}
This work is supported by the DST (India) fast track project (grant no:SR/FTP/PS-74/2008). TM acknowledges T. Nautiyal for useful discussions. SK acknowledges 
CSIR (India) for a research fellowship. 


\end{document}